\documentclass[aps,prd,reprint,showpacs,article,floatfix]{revtex4-1}
\pdfoutput=1 

\usepackage{fullpage}
\usepackage{amsmath}
\usepackage{amssymb}
\usepackage{graphicx}
\usepackage{xcolor}
\usepackage{hyperref}

\begin{document}

\title{A Microscopic Description of Thermodynamic Volume\\  in\\ Extended Black Hole Thermodynamics}

\author{Clifford V. Johnson}
\email{johnson1@usc.edu}
\affiliation{Department of Physics and Astronomy\\ University of
Southern California \\
 Los Angeles, CA 90089-0484, U.S.A.}

\author{Victoria L. Martin}
\email{victoria.martin.2@asu.edu}
\affiliation{Department of Physics\\ Arizona State University\\
Tempe, AZ 85287-1504, U.S.A.}

\author{Andrew Svesko}
\email{andrew.svesko@asu.edu}
\affiliation{Department of Physics\\ Arizona State University\\
Tempe, AZ 85287-1504, U.S.A.}

%\pacs{05.70.Ce,05.70.Fh,04.70.Dy}

\begin{abstract}
Using the fact that theories of gravity with asymptotically three--dimensional anti--de Sitter  geometries have dual descriptions as  two--dimensional conformal field theories (CFTs), we present the first study  in field theory of  the thermodynamic volume of various black hole solutions. We  explain in general two--dimensional CFT terms why the presence of a thermodynamic volume can render certain black hole solutions ``super--entropic''. Super--entropicity simply results from the fact that the Cardy formula, which gives the gravitational Bekenstein--Hawking entropy,  can over--count the CFT entropy. %, assuming only positivity of the volume. 
The examples of charged Ba\~nados, Teitelbiom and Zanelli (BTZ) black holes and generalized exotic BTZ black holes are described.  These observations help explain why  the specific heat at constant volume can signal the instability of such solutions, as recently conjectured.  
\end{abstract}


\maketitle

\section{Introduction}
A decade ago,  a new variable was introduced into  black hole thermodynamics, as part of an extended  thermodynamic framework~\cite{Kastor:2009wy}. It has the interpretation as the thermodynamic volume, $V$, dual to the pressure $p\,{\equiv}{-}\Lambda/8\pi$ that is present if the cosmological constant $\Lambda$ is  dynamical~\footnote{This paper will not be concerned with the origin of the dynamics of $\Lambda$. It can be achieved in various ways if the gravity sector is embedded within a larger framework. An example is when there are also dynamical scalars $\varphi_i$ with a potential $V(\varphi_i)$. Fixed points of the potential, where the scalars take on fixed values $\varphi^{\rm c}_i$, gives a non--zero constant $V(\varphi^{\rm c}_i)$,  determining the value of $\Lambda$ in anti--de Sitter vacua of the theory. See {\it e.g.} the review  in ref.~\cite{Aharony:1999ti}.}. The pair $(p,V)$ joins the variables of traditional gravitational thermodynamics~\cite{Bekenstein:1973ur,Bekenstein:1974ax,Hawking:1974sw,Hawking:1976de}\footnote{For a recent historical overview, see {\it e.g.,} ref.~\cite{Grumiller:2014qma}.}, the entropy $S{=}A/4$ and temperature $T{=}\kappa/2\pi$, associated to a black hole horizon's area $A$, and surface gravity $\kappa$. The black hole's mass~$M$ becomes identified with the enthalpy $H{=}U{+}pV$, where~$U$ is the internal energy. (Here, we are using geometrical units where $G,c,\hbar,k_{\rm B}$ have been set to unity.) The first law of thermodynamics for black holes then becomes:
\begin{equation}
\label{eq:first-law}
dH=TdS+Vdp\ ,
\end{equation}
(assuming for the moment that no other dynamical quantities, like charge and angular momentum, are in play).
 The volume is a derived quantity, calculated after the mass (hence enthalpy $H(S,p)$) of the solution has been identified: $V{\equiv}\left.\partial H/\partial p\right|_S$ . In the simple case of static black holes (with no additional non--trivial scalar sector) it has the geometric interpretation as the naive spherical volume occupied by the black hole, but in general it is non--geometrical~\cite{Cvetic:2010jb,Johnson:2014xza}. In such general settings, it becomes a truly independent variable from the entropy, and the physics associated with it becomes richer. 

In the context of negative $\Lambda$,
%(so positive $p$, and hence a  interpretable thermodynamics),
 where the gravitational physics can often be recast in terms of a dual (non--gravitational) field theory in one dimension fewer using the correspondence between anti--de Sitter dynamics and conformal field theory physics (the AdS/CFT correspondence)~\cite{Maldacena:1997re,Gubser:1998bc,Witten:1998qj,Witten:1998zw}, it is natural to ask whether~$V$ has a direct interpretation in the field theory~\footnote{See refs.~\cite{Kastor:2009wy,Dolan:2013dga,Johnson:2014yja} for early ideas and remarks, and refs.~\cite{Dolan:2014cja,Couch:2016exn} for some explorations.}. In general, this question is rather hard to explore, since the duality addresses the strongly coupled field theory regime, which is not always easily accessible in traditional field theory terms. Moreover the finite $T$ regime of the AdS/CFT duality is (in general)  rather less well robustly explored than the $T{=}0$ sector. 

In the present work, we point out that progress can be made in the case of three dimensional gravity (with $\Lambda{<}0$), since in that case the duality's dictionary is rather stronger: Asymptotically anti--de Sitter geometries in three dimensions (AdS$_3$) are dual to conformally invariant two dimensional  field theories, which are very tightly constrained in their structure.  Moreover, the finite temperature~$T$ is simply the (inverse) period of a cycle in the two dimensional  Riemann surface the theory is defined on. We will be able to write (in sections~\ref{sec:cft-btz} and~\ref{sec:charged-btz}) the thermodynamic volume $V$ in terms of quantities very familiar in the  CFT. With that achieved, it is then straightforward to translate any conditions involving~$V$ into statements in the CFT. 

For example, it is natural in thermodynamics to ask questions about the fixed volume sector. However, in general \footnote{For static black holes with no scalars, $V$ and $S$ are not independent and so in those simple cases fixed~$V$ is simply fixed area. For most cases however, $V$ is a non--geometrical quantity independent of $S$.}, this is  somewhat mysterious from the black  hole thermodynamics perspective---fixed pressure is more natural there since that is simply fixed~$\Lambda$---but with a microscopic dual field theory identification such  as the one presented here, progress can be made in examining the physics of the fixed volume sector. (This may be of use in furthering  recent work~\cite{Johnson:2019vqf,Johnson:2019olt,Johnson:2019mdp} that has uncovered novel and potentially useful physics in the fixed volume sector of black hole thermodynamics.)

We make such progress by  arriving, in an important example, at a microscopic connection between the thermodynamics of the fixed volume sector and a phenomenon called ``super--entropicity'' identified for a class of black holes~\cite{Cvetic:2010jb}. These are black holes with volume~$V$ that have an  entropy $S$ that exceeds the amount that a Schwarzschild black hole with the same volume would have. It was noticed in ref.~\cite{Johnson:2019mdp} that several solutions for which this is the case have a negative specific heat at constant volume, $C_V$, signalling an instability. For one example, the charged BTZ solution, this could be demonstrated analytically, and a connection between the super--entropicity and the negativity emerged (reviewed below). This led to a more general conjecture that super--entropicity implies the $C_V<0$ instability. 

While we do not prove the conjecture here, we find a microscopic phenomenon that seems to explain (or at least herald) the super--entropicity on the gravity side, and it emerges precisely as a result of our microscopic identification of the thermodynamic volume~$V$ and as a  consequence of working in the fixed $V$ sector! It works as follows: The  standard (microscopic) CFT expression for the entropy, $S$, of the black holes  which successfully reproduces \cite{Carlip:1994gc,Strominger:1996sh,Strominger:1997eq,Birmingham:1998jt}  the gravitational Bekenstein--Hawking entropy, is usually the Cardy formula \cite{Cardy:1986ie,Bloete:1986qm} in these dualities, and it turns out to be built out of some of the same quantities as the thermodynamic volume~$V$. What we show is that working at fixed, positive~$V$ places a condition on the CFT sector meaning that the (naive) Cardy formula {\it over--counts} the entropy in the CFT. This is the microscopic herald of the fact that the gravity entropy (as counted by Cardy) is, in a precise sense, ``too much''. This is presented in detail in sections~\ref{sec:charged-btz} and~\ref{sec:super-entropic}.

As a simple first check of our microscopic formalism and our assertion that super--entropicity is connected to the over--counting seen in the CFT, we study a rather large family of examples in section~\ref{sec:exotic-btz}. These ``generalized exotic'' BTZ black holes~\cite{Carlip:1991zk,Carlip:1994hq,Townsend:2013ela} have a rich extended thermodynamics~\cite{Cong:2019bud} with $C_V{\neq}0$ that can also be written in  two dimensional  CFT terms. While there are sectors that have negative specific heats (both $C_p$ and $C_V$ can be negative for some ranges of parameters, and positive for other ranges) these examples, which are non--unitary in some cases,  are {\it not} super--entropic~\footnote{Here we disagree with the interpretation of  ref. \cite{Cong:2019bud}. They use a definition of super--entropic inherited from the geometrical formula of ref.~\cite{Cvetic:2010jb} that focuses on area $A$, and not entropy, $S$. They therefore conclude that there is a problem with the conjecture connecting super--entropicity to negative $C_V$ since they can find regions with positive $C_V$.  However, we are (as is ref.~\cite{Johnson:2019mdp}) using the entropy--focused interpretation of the term super--entropic as opposed to the (less physical) area--focused usage. See sections~\ref{sec:super-entropic} and~\ref{sec:exotic-btz} for more discussion.}. In the spirit of our methods,  the thermodynamic volume~$V$ can be written in terms of CFT quantities. Doing  so, we see that working at fixed $V$  does {\it not} result in the Cardy formula over--counting the CFT entropy. This therefore fits with our suggestion that super--entropicity is heralded by such an over--count at fixed $V$.

In section~\ref{sec:conclusion} we discuss the results further, including ideas and prospects for extending this microscopic success to higher dimensions.

\section{CFT and Standard BTZ}
\label{sec:cft-btz}

Two principal quantities in  two dimensional conformal field theory are the energy $E$ and the spin $J$, which  are given in terms of the sum and difference of the eigenvalues, $\Delta,{\bar \Delta}$,  of $L_0$ and ${\bar L}_0$, the zeroth components of the right and left Virasoro generators (which define the conformal algebra):
\begin{equation}
\label{eq:energy-spin}
E = \frac{\Delta+{\bar \Delta}}{\ell}\ , \quad J=\Delta-{\bar \Delta}\ .
\end{equation}
Here $\ell$ is a length scale set by the cosmological constant of the dual gravity theory {\it via} $\Lambda=-1/\ell^2$. The right and left Virasoro algebras have central charges~$c_R$ and $c_L$, which are proportional to $\ell$. Their precise values are example dependent, as we shall see. The values of $E$ and $J$ are computed in the dual gravity theory quite readily, and are the mass~$M$ and angular momentum~$J$ of the black hole spacetime. The entropy on the gravitational side is computed using the Bekenstein--Hawking formula, the quarter of the area of a horizon.  (Note that ``area'' here will mean the circumference of a circle, since there are only two spatial dimensions in the gravity theory. There may be contributions from more than one horizon, as we shall see in later examples.) On the field theory side, this entropy is reproduced in the field theory using~\cite{Carlip:1994gc,Strominger:1996sh,Strominger:1997eq,Birmingham:1998jt} the Cardy formula for the asymptotic degeneracy of states with a given conformal dimension:
\begin{equation}
\label{eq:cardy}
S = \log(\rho(\Delta,{\bar\Delta}))=2\pi\sqrt{\frac{c_R\Delta}{6}}+2\pi\sqrt{\frac{c_L{\bar\Delta}}{6}}\ .
\end{equation}
Crucially, this formula's validity depends upon the key  assumption that the lowest  $L_0,{\bar L}_0$  eigenvalues vanish~\cite{Carlip:1998qw}. We will revisit this issue shortly.

For  the examples discussed in this paper, the spacetime metric will be of the leading Ba\~nados, Teitelbiom and Zanelli (BTZ)~\cite{Banados:1992wn,Banados:1992gq} form:
\begin{eqnarray}
\label{eq:black-hole}
ds^2 &=& -f( r)dt^2
+ f(r)^{-1}dr^2 + r^2 \left(d\varphi-\frac{4j}{r^2}dt\right)^2\ ,\nonumber\\
{\rm with} && f( r) = -8m+\frac{r^2}{\ell^2}+\frac{16j^2}{r^2}+\cdots\ , \label{eq:metric}
\end{eqnarray}
(with one exception we will discuss separately).
The black hole has an outer and inner horizon, at radii denoted $r_{\pm}$, which are the larger and smaller roots of $f(r)=0$. Depending upon the parent gravity theory in question (examples below), the parameters $m$ and $j$ determine the black hole mass $M$ and angular momentum $J$ either directly or in linear combination. The classic  BTZ example has $f(r)$ as written (no extra terms) and $M{=}m$ and $J{=}j$, and together with $S$ they are:
\begin{equation}
\label{eq:gravity-btz}
M=\frac{r_+^2+r_-^2}{8\ell^2}\ ,\quad J=\frac{r_+r_-}{4\ell}\ ,\quad S=\frac{\pi r_+}{2}.
\end{equation}
Comparing the first two quantities to those in equation~(\ref{eq:energy-spin}) gives, after a little algebra:
\begin{equation}
\label{eq:Deltas}
\Delta = \frac{(r_++r_-)^2}{16\ell}\ , \quad{\rm and}\quad{\bar\Delta} = \frac{(r_+-r_-)^2}{16\ell}\ .
\end{equation}
Using these in equation~(\ref{eq:cardy}) with $c_R{=}c_L{=}3\ell/2$ yields the gravity entropy in equation~(\ref{eq:gravity-btz}). 

In extended thermodynamics, the pressure is given by $p{=}1/8\pi\ell^2$, and the mass $M$ is the enthalpy 
\begin{equation}
\label{eq:enthalpy}
H(S,p)=4\pi p \left(\frac{S}{\pi}\right)^2+\frac{\pi^2J^2}{2S^2}\ .
\end{equation}
We will work at fixed $J$ henceforth, treating it as a parameter. The first law remains as in equation~(\ref{eq:first-law}).
Hence, the thermodynamic volume and temperature turn out to be 
\begin{equation}
\label{eq:temp-vol}
V\equiv\left.\frac{\partial H}{\partial p}\right|_S=\pi r_+^2\ ,\quad%{\rm and}\,\,\,
 T\equiv\left.\frac{\partial H}{\partial S}\right|_p=\frac{r_+^2-r_-^2}{2\pi\ell^2 r_+}\ ,
\end{equation}
the latter agreeing with either a surface gravity computation or the requirement of regularity of the Euclidean section~\cite{Gibbons:1976ue}.

We can go a step further. The CFT/gravity relations~(\ref{eq:Deltas}) can be inverted to give $r_\pm$ in terms of $\Delta$ and ${\bar \Delta}$, and so we can write $V$ in terms of CFT quantities as:
 \begin{equation}
 \label{eq:volume-microscopic}
 V=\frac{8\pi}{3}\left(\sqrt{c_R\Delta}+\sqrt{c_L{\bar\Delta}}\right)^2\ .
 \end{equation}
 We propose that this relationship should be read in an analogous manner to how the Cardy formula in equation~(\ref{eq:cardy}) is read. States can be constructed in the CFT in the usual manner, acting on the vacuum with the left and right (negatively moded) Virasoro generators as creation operators. Then $L_0$ and ${\bar L}_0$ measure $\Delta$ and ${\bar \Delta}$. For given values of these quantities, equation~(\ref{eq:volume-microscopic}) defines a quantity $V$ that has the interpretation as the thermodynamic value in the gravity theory. Since it is made from (the square of) the same combination of CFT quantities that $S$ is built from, there is not much more to learn from this example. Questions about $V$ are equivalent to questions about~$S$, as they are not independent quantities.

%%%%%%%%%%%%%%%%%%%%%%
%%%%%%%%%%%%%%%%%%%%%%

\section{Charged BTZ Black Holes}
\label{sec:charged-btz}

Our first example where something new arises is the charged BTZ black hole with no angular momentum, a solution of Einstein--Maxwell in three dimensions~\cite{Martinez:1999qi}. Now, we have $J{=}0$ and  the metric function to use in equation~(\ref{eq:metric}) is instead $f(r){=}-8M+\frac{Q^2}{2}\log\left({r}/{\ell}\right)+{r^2}/{\ell^2}$, where~$Q$ is the $U(1)$ charge of the solution and $M$ is the mass. There is also a gauge field $A_t = Q\log\left({r}/{\ell}\right)$. From the point of view of the two dimensional CFT, $Q$ is merely a deformation parameter, a global charge, which will be kept fixed here. The extended thermodynamics gives~\cite{Frassino:2015oca}:
\begin{eqnarray}
\label{eq:thermodynamic-quantities}
H&=&\frac{4pS^2}{\pi} - \frac{Q^2}{32}\log\left(\frac{32pS^2}{\pi}\right)\ ,\quad S=\frac{\pi}{2}r_+\ ,\nonumber\\
T&=&\frac{8pS}{\pi}-\frac{Q^2}{16S}\ ,\quad {\rm and}\quad V= \frac{4S^2}{\pi} -\frac{Q^2}{32p}\ ,
\end{eqnarray}
and the first law is again equation~(\ref{eq:first-law}). The internal energy of the system is given by $U{\equiv}H{-}pV{=}(Q^2/32)[1-\log(32pS^2/\pi)]$. 

Note that the presence of the charge  $Q$ introduces a $\log(r/\ell)$ term in the metric function $f(r)$. Consequently, the asymptotic symmetry group of the geometry is deformed, hiding the action~\cite{Brown:1986nw} of the Virasoro algebra. Crucially, we regard Virasoro as hidden, but not absent. We propose that the conformal field theory will still have the structure that we saw in the previous example, and below we will find strong evidence in support of this.

 To make Virasoro explicit requires a different approach. The boundary conditions on the metric and gauge field can be modified by enclosing the entire black hole system inside some radius~$r_{0}$ and introducing a renormalized mass according to $M(r_{0})=M+\frac{Q^{2}}{16}\log\left(r_{0}/\ell\right)$, such that the manifest asymptotic Virasoro symmetry  is restored \cite{Cadoni:2007ck}.  This alternative scheme rearranges the thermodynamic quantities (both traditional and extended). In the resulting extended thermodynamics (which requires promoting the scale $r_0$ to a dynamical variable in order to have a consistent first law~\cite{Frassino:2015oca}) the thermodynamic volume $V$ loses its $Q$ dependence, becoming the geometric volume $\pi r_+^2$, and since $S{=}\pi r_+/2$, we have  $C_{V}{=}0$. Hence,  we will not study this renormalized scheme %(where there is no super--entropicty), 
 and instead focus our attention on the thermodynamic quantities as presented in equations~(\ref{eq:thermodynamic-quantities}), which yield an interesting case study.  We will revisit the issue of the renormalized scheme in later discussion.

Notice that~$V$ and $S$ in equation~(\ref{eq:thermodynamic-quantities}) are now independent. The requirement that the temperature be positive results in the restriction $Q^2{\leq}4\eta$, where $\eta{=}32pS^2/\pi$. Since $V{=}TS/2p$, this also translates into positivity of the volume $V$. The parameter $\eta$ also appears in the internal energy $U$, and requiring that $U{>}0$ gives $\eta{\leq}1$. So, just from the gravity side, we get the bound~$Q^2{\leq}4$.

Turning to the  CFT quantities, $c_L{=}c_R{=}3\ell/2{=}c$ as before, and since $J{=}0$ we have  $\Delta{=}{\bar\Delta}$. The Cardy formula gives the entropy as before: $S=4\pi\sqrt{c\Delta/6}$, but now the thermodynamic volume $V$, written in terms of CFT quantities, is:
\begin{equation}
\label{eq:thermodynamic-volume-charged-btz}
V=\frac{32\pi c}{3}\left(\Delta-\frac{Q^2c}{96}\right)\ .
\end{equation} 
Positivity of $V$ (following from positivity of $T$) translates into a non--trivial statement: The lowest $\Delta$ can be is $\Delta_0{=}Q^2c/96$. Recall that an assumption underlying the Cardy formula~(\ref{eq:cardy}) is that $\Delta_0{=}0$. In fact, when $\Delta_0{\neq}0$, the correct formula to use for the (logarithm of the) asymptotic density of states replaces $c$ by $c_{\rm eff}{\equiv}c{-}24\Delta_0$, resulting in (for positive $\Delta_0$) a reduction of the entropy count \cite{Carlip:1998qw}. For us, $c_{\rm eff}{=}c(1{-}Q^2/4)$, and we recover two interesting pieces of information. The first is that the gravity entropy, which corresponds to the naive Cardy formula, {\it over--counts the number of degrees of freedom of the theory}. The second is that there is a unitarity bound  of $Q^2{\leq}4$, the same bound we obtained by independent gravity requirements that~$T$ and $U$ are positive!

That we have recovered precisely the same condition on $Q$ using two very different considerations (gravity and CFT) is strong support for our proposal for writing a microscopic/CFT formula for $V$. It also strongly suggests that we were correct to  use  the AdS$_3$/CFT$_2$ map for this charged black hole despite the fact that the asymptotic algebra is deformed by the presence of~$Q$.

The  over--counting of the entropy discovered here suggests that something is wrong with the equilibrium thermodynamics suggested by the variables in equation~(\ref{eq:thermodynamic-quantities}). We propose that  it is in fact  a herald of the phenomenon called ``super--entropicity'', discussed next.

\section{Super--Entropicity and Instability}
\label{sec:super-entropic}

The charged BTZ solution is the simplest example of a super--entropic black hole \cite{Frassino:2015oca}. Generally, a super--entropic $d$--dimensional black hole is defined as a solution which violates the reverse isoperimetric inequality \cite{Cvetic:2010jb}:
\begin{equation}
 \mathcal{R}\equiv\left(\frac{(d-1)V}{\omega_{d-2}}\right)^{\frac{1}{d-1}}\left(\frac{\omega_{d-2}}{4S}\right)^{\frac{1}{d-2}}\geq1\;,\label{revisogend}
\end{equation}
where $V$ is the thermodynamic volume, and~$S$ is the gravitational entropy. Also, the quantity $\omega_{n}{=}2\pi^{(n+1)/2}/\Gamma[(n+1)/2]$ is the standard volume of the round unit sphere. The inequality (\ref{revisogend}) is saturated by  Schwarzschild--AdS black holes (BTZ in $d{=}3$),  with 
$\mathcal{R}{=}1$. Black holes where $\mathcal{R}{>}1$ are said to be sub--entropic, such as Kerr--AdS \cite{Cvetic:2010jb} and STU black holes \cite{Caceres:2015vsa}. 
Systems with $\mathcal{R}{<}1$, such as the ultra--spinning limit of Kerr--AdS black holes \cite{Hennigar:2014cfa,Hennigar:2015cja}, are super--entropic. In $d{=}3$, the rotating BTZ black hole has $\mathcal{R}{=}1$, while the charged BTZ hole has $\mathcal{R}{<}1$. 

Notice that the inequality (\ref{revisogend}) is written here with~$\mathcal{R}$ defined in terms of the entropy $S$ instead of the horizon area~$A$, as it is  written in  ref.\,\cite{Cvetic:2010jb}. This is because, in our view,  super--entropicity is a statement about the thermodynamic quantity {\it entropy} (as the title suggests) and not about the outer horizon area~\footnote{A similar modification to the reverse isoperimetric inequality was made in ref.~\cite{Feng:2017jub} for black hole solutions of Horndeski theories of gravity.}. Moreover, more general theories of gravity have an entropy that is not proportional to the outer horizon area, but may include contributions from the inner horizon~\footnote{In fact, sometimes even in ordinary gravity,  the entropy receives contributions from other objects. See the Taub--NUT and Taub--Bolt examples in refs.~\cite{Chamblin:1998pz,Emparan:1999pm,Mann:1999pc}.}.  The next section will have such examples, and as we will see, our definition (\ref{revisogend}) will lead to a different interpretation of their entropic character than one proposed in the recent literature~\cite{Cong:2019bud}. 

It was  recently observed \cite{Johnson:2019mdp} that several super--entropic black holes are thermodynamically unstable, signified by a negative heat capacity $C_{V}$. It was conjectured there that  super--entropicity may generally imply that $C_V{<}0$, following from the fact that for  a charged BTZ black hole this can be verified analytically: The temperature $T$ and $C_{V}$ take the form:
\begin{eqnarray}
 T&=&\frac{\pi V}{16S}\frac{Q^{2}}{(4S^{2}-\pi V)},\; \nonumber \\
 C_{V}&=&-S\left(\frac{4S^{2}-\pi V}{12S^{2}-\pi V}\right)\;.\label{TCVchargedbtz}
\end{eqnarray}
The temperature is positive when $4S^{2}{>}\pi V$, which is equivalent to the $d{=}3$ super--entropicity condition~$\mathcal{R}{<}1$. Moreover, this is precisely when the charged BTZ solution has  $C_{V}{<}0$, {\it i.e.,} it is thermodynamically unstable.
(Showing that $C_{V}{<}0$ when~$\mathcal{R}{<}1$   was also verified numerically in ref.~\cite{Johnson:2019mdp} for a class of  ultra--spinning Kerr--AdS black holes in various higher dimensions. Analytic counterparts to the above $d{=}3$ demonstration were not obtained however.) 

Positivity of $T$ ensuring a connection between super--entropicity and instability is strongly reminiscent of what we saw in the previous section, when making connections to the CFT.  When the dual CFT is unitary, we may translate $c_{\text{eff}}{>}0$ into $4S_{\text{CFT}}^{2}{>}\pi V$, where  $S_{\text{CFT}}{=}4\pi\sqrt{c_{\text{eff}}\Delta/6}$. Then, since $S{>}S_{\text{CFT}}$, we have $4S^{2}{>}\pi V$. Therefore, super--entropicity reflects that the gravitational entropy over--counts the number of degrees of freedom of the underlying microscopic theory. 

The over--counting is also accompanied by the negativity of $C_{V}$, which itself suggests an instability, a movement in solution space to some new set of thermodynamic quantities for which $C_V$ is no longer negative. It is tempting to speculate that the extended thermodynamics yielded~\cite{Frassino:2015oca} by studying  the renormalized scheme of ref.~\cite{Cadoni:2007ck} (reviewed briefly below equations~(\ref{eq:thermodynamic-quantities})) is the endpoint of the instability. One suggestion of our observations here is that there is another framework (different from the renormalization scheme  recalled below equation~(\ref{eq:thermodynamic-quantities})) in which the asymptotic Virasoro algebra is restored, but in which the central charge is modified to our effective central charge $c_{\rm eff}{=}c(1{-}Q^2/4)$. It would be interesting to find such a framework, and to see whether the resulting thermodynamic quantities produce a super-- or sub--entropic system.

%%%%%%%%%%%%%%%%%%%%%%%%%%%%%
%%%%%%%%%%%%%%%%%%%%%%%%%%%%%

\section{Generalized Exotic BTZ \\Black Holes}
\label{sec:exotic-btz}

As a final example we consider the family of ``generalized exotic BTZ'' black holes \cite{Carlip:1991zk,Carlip:1994hq,Townsend:2013ela}. The relevant gravity theory is a linear combination of the Einstein--Hilbert action and the gravitating Chern--Simons action, $I=\alpha I_{\rm EM}+\gamma I_{\rm GCS}$, where $\gamma=1{-}\alpha$. The metric is again given in equation~(\ref{eq:black-hole}), with no extra terms for $f(r)$, but this time the mass and angular momentum mix the parameters $m$ and $j$: $M=\alpha m{+}\gamma j/\ell$, $J=\alpha j {+} \gamma \ell m$. The case of $\alpha{=}1$ is the standard BTZ black hole, while $\gamma{=}1$ is the exotic BTZ black hole. General $0\leq\alpha\leq1$ interpolates between these two extremes. The thermodynamic variables are given by:
\begin{eqnarray}
\label{eq:thermodynamics-exotic}
M&=&\frac{\alpha(r_+^2+r_-^2)}{8\ell^2} + \frac{\gamma r_+r_-}{4\ell^2}\ , \nonumber \\
J&=&\frac{\alpha r_+r_-}{4\ell}+\frac{\gamma(r_+^2+r_-^2)}{8\ell}\ , \nonumber \\
T&=&\frac{r_+^2-r_-^2}{2\pi\ell^2 r_+}\ ,\quad \Omega=\frac{r_-}{r_+\ell}\ , \nonumber \\
S&=&\frac{\pi}{2}(\alpha r_++\gamma r_-)\ , \nonumber\\ 
V&=&\alpha\pi r_+^2+\gamma\pi r_-^2\left(\frac{3r_+}{2r_-}-\frac{r_-}{2r_+}\right)\ ,
\end{eqnarray}
where $\Omega$ is the angular velocity. 

Recently it was shown that generalized exotic BTZ solutions can have $C_{V}$ both positive and negative \cite{Cong:2019bud}. Specifically, for $\alpha<1/2$, $C_{V}$ is positive for large enough $r_{+}$. In the regions where $C_{V}>0$, however, the heat capacity at constant pressure $C_{p}$ will be negative, indicating that they are generally unstable. Notice that for the  inequality (\ref{revisogend}), we have 
\begin{equation}
\mathcal{R}=\frac{1}{2(\alpha+\gamma x)}\sqrt{4\alpha+6\gamma x-2\gamma x^{3}}\;\ ,
\end{equation}
where $x\equiv r_{-}/r_{+}$ ranges between $0$ and $1$. For the defined range of non--zero $\alpha$, we find $\mathcal{R}>1$, and thus these generalized exotic BTZ black holes form a class of {\it sub--entropic} black holes. Had we instead used the form of $\mathcal{R}$ first written  in \cite{Cvetic:2010jb}, we would have found $\mathcal{R}<1$  and concluded that these solutions are super--entropic, as ref. \cite{Cong:2019bud} does~\footnote{We thank W. Cong and R. B. Mann for helpful communications about this matter.}. However, as we have already stated, we are using the entropy--focused interpretation of the term super--entropic as opposed to the (less physical) area--focused usage. In this sense, in the spirit of ref.~\cite{Johnson:2019mdp}'s conjecture and what we've seen in the previous two sections, there is no super--entropicity and hence $C_V$ does not need to become negative, since the solution does not need to somehow shed the extra entropy.

Turning to the dual conformal field theory, some algebra shows that variables $M$, $J$, and $S$  fit the CFT form given in equations~(\ref{eq:energy-spin}) and~(\ref{eq:cardy}), (with    factors  $\alpha{+}\gamma{=}1$ for  right--moving quantities  and $\alpha{-}\gamma{=}2\alpha{-}1$  for left--moving):
\begin{eqnarray}
\Delta &=& \frac{1}{16\ell}(r_+^2+r_-^2)\ ,\quad
{\bar \Delta} = \frac{2\alpha-1}{16\ell}(r_+^2-r_-^2)\ , \nonumber\\
c_R &=& \frac{3\ell}{2}\ ,\quad c_L  =  \frac{3\ell}{2} (2\alpha-1)\ .
\end{eqnarray}
%Notice that the cases $\alpha<1/2$ are non--unitary. 
We may recast the thermodynamic volume $V$ (\ref{eq:thermodynamics-exotic}) in terms of these CFT parameters. The resulting expression is:
\begin{eqnarray}
\label{eq:volexoticcft}
& \frac{3V}{4\pi c_{R}}&=\left(1+\frac{1}{\epsilon}\right)\left(\sqrt{\Delta}+\sqrt{\epsilon\bar{\Delta}}\right)^{2} \\
&+&\left(1-\frac{1}{\epsilon}\right)\left(\frac{\sqrt{\Delta}-\sqrt{\epsilon\bar{\Delta}}}{\sqrt{\Delta}+\sqrt{\epsilon\bar{\Delta}}}\right)\left[\Delta+\epsilon\bar{\Delta}+4\sqrt{\epsilon\Delta\bar{\Delta}}\right]\;,\nonumber
\end{eqnarray}
where %we have defined
%\begin{equation} 
$\epsilon{\equiv}{c_{R}}/{c_{L}}$.
%\end{equation}
% long and not needed here, so it won't be displayed.
Note that  $c_R{=}c_L$ when $\alpha{=}1, \gamma=0$, {\it i.e.}, we have the usual BTZ solution of section~\ref{sec:cft-btz}, and our expression  (\ref{eq:volexoticcft}) reduces to the thermodynamic volume given in equation~(\ref{eq:volume-microscopic}). 
%When $\alpha\neq1$, $V$ is the thermodynamic volume in the generalized exotic BTZ solution. 

The key observation from (\ref{eq:volexoticcft}) is that, unlike the charged BTZ case,  requiring  positivity of~$V$ does not lead to a shift away from zero for  the lowest value of $\Delta$ or $\bar{\Delta}$. As such, the gravitational entropy (as given by the Cardy formula) does not over--count the number of microscopic degrees of freedom.  This fits with the observation above that there is no super--entropicity in these examples (using the entropy--focused definition of ${\cal R}$ in equation~(\ref{revisogend})).

\section{Conclusion}
\label{sec:conclusion}
In conclusion, we have shown how to microscopically interpret (using AdS$_3$/CFT$_2$ duality) the thermodynamic volume of extended black hole thermodynamics, by writing formulae for it in terms of CFT quantities. For simple black holes where $V$ and $S$ are not independent, such a formula is no more useful than the Cardy formula for $S$. However, deploying the interpretation in the charged BTZ example where $V$ is independent of $S$, we  uncovered that the naive Cardy formula over--counts the entropy of the theory. We interpret this as a microscopic herald  of the super--entropicity phenomenon associated to some solutions in extended thermodynamics.

Independent conditions derived from gravity and CFT gave precisely the same bound on $Q$, the black hole charge: $Q^2{\leq}4$, suggesting  internal consistency of our methods. These methods included using the CFT dual of the charged black hole solution even though the presence of $Q$ deforms the asymptotic symmetry (Virasoro) algebra. This might suggest that there  is another framework (different from the renormalization scheme  recalled below equation~(\ref{eq:thermodynamic-quantities})) in which the asymptotic Virasoro algebra is restored, but in which the central charge is modified to our effective central charge $c_{\rm eff}{=}c(1{-}Q^2/4)$. It would be interesting to find such a framework, and to see whether the resulting thermodynamic quantities produce a super-- or sub--entropic system.

It would  also be interesting to find a similar microscopic understanding of super--entropicity of ultra--spinning black holes~\cite{Hennigar:2014cfa,Hennigar:2015cja}. These solutions exist for $d\,{\geq}\,4$, where we can no longer use $\text{AdS}_{3}/\text{CFT}_{2}$ duality. Instead, however, we could consider Kerr/CFT duality \cite{Guica:2008mu}, along the lines of ref.~\cite{Sinamuli:2015drn}, and see if constraints imposed by the gravitational thermodynamics lead to any requirements on the dual CFT. We leave this  for future work. 

Another line of investigation could be to develop further a  characterization of how super--entropicity may result in the $C_V{<}0$ instability for other black holes, and in other dimensions.  As conjectured in ref.~\cite{Johnson:2019mdp}, a consequence  of super--entropicity is  negativity of $C_{V}$. (Note again that this is not the same as saying that negativity of $C_V$ implies super--entropicity.) For the charged BTZ case this was shown directly in equation~(\ref{TCVchargedbtz}), where the form and sign of $C_{V}$ depends solely on the ability to write the temperature as
$T{=}{\mathcal{F}(S,V,Q)}/{(1{-}\mathcal{R})}\ ,$
where $\mathcal{F}$ is a function we wish to characterize further, and the $\mathcal{R}$ in the denominator is  given in equation~(\ref{revisogend}).
 Not every black hole solution will have a temperature that can be written in this form, as we see in the cases of the  uncharged and exotic BTZ black holes. Moreover, we know of sub--entropic solutions whose temperature does take this form, {\it e.g.}, the $d{=}4$ Kerr--AdS black hole \cite{Dolan:2011xt}. 
Nonetheless, we might attempt to learn something about  a sub--class of super--entropic black hole solutions by demanding the temperature take the form given above. If they have negative $C_{V}$, it implies conditions on $\mathcal{F}$. Our special form of $T$ together with the fact that $T{=}f'(r_{+})/{4\pi}$ for a gravity solution with metric function $f(r)$  might characterize enough about the properties of  $f(r)$ to use it as a diagnostic tool for a wide variety of solutions. We leave this  for  future work.

%\medskip
 
 \begin{acknowledgments}
CVJ's work was supported by the  US Department of Energy under grant  DE-SC0011687. Some of this work was carried out at the Aspen Center for Physics, which is supported by National Science Foundation grant PHY-1607611.  The work of VLM  is supported by the U.S. Department of Energy under grant number DE-SC0018330. AS and VLM thank the USC Department of Physics and Astronomy, where work on this project began, for hospitality. CVJ thanks Amelia for her support and patience.    
\end{acknowledgments}

\bibliographystyle{apsrev4-1}
\bibliography{btz_cft_volume}

\end{document}